\documentclass[letterpaper, 10 pt, conference]{ieeeconf}
\IEEEoverridecommandlockouts 
\title{Adaptive model predictive control for traffic signal timing with unknown demand and parameters}

\author{Zhexian Li \qquad Ketan Savla 
		\thanks{The authors are with the Sonny Astani Department of Civil and Environmental Engineering, University of Southern California, Los Angeles, CA. 
		\texttt{\{zhexianl,ksavla\}@usc.edu}.
		This work was supported in part by METRANS 23-07. 
		K. Savla has a financial interest in Xtelligent, Inc.}}

\usepackage{amsmath,amssymb,bbm,color,psfrag,amsfonts,mathtools}
\usepackage{dsfont}
\usepackage{graphicx,cuted}                       

\usepackage{wrapfig}
\usepackage{pdfpages}
\usepackage{accents}
\usepackage{lipsum}
\usepackage[T1]{fontenc}
\usepackage[utf8]{inputenc}
\usepackage{mathtools}
\usepackage{multicol}
\usepackage{paralist}   
\usepackage{figlatex}
\let\labelindent\relax
\usepackage{enumitem}
\usepackage{booktabs}  
\usepackage{multirow}
\usepackage[normalem]{ulem}

\usepackage{caption}
\usepackage{subcaption}

\usepackage{cite}
\usepackage[pdftex,pdfpagelabels,bookmarks,hyperindex,hyperfigures]{hyperref}

\usepackage{verbatim}
\usepackage{algorithm}
\usepackage{algorithmic}

\newcounter{problem}

\makeatletter
\newcommand\footnoteref[1]{\protected@xdef\@thefnmark{\ref{#1}}\@footnotemark}
\makeatother

\newtheorem{theorem}{Theorem}

\newtheorem{lemma}{Lemma}

\newtheorem{definition}{Definition}
\newtheorem{assumption}{Assumption}

\newtheorem{remark}{Remark}
\newtheorem{example}{Example}

\usepackage{color}
\newcommand{\Link}{\mathcal{L}}
\newcommand{\Lint}{\mathcal{L}_{\text{int}}}
\newcommand{\Lentry}{\mathcal{L}_{\text{entry}}}
\newcommand{\Lout}{\mathcal{L}^{\text{out}}}
\newcommand{\Lexit}{\mathcal{L}_{\text{exit}}}
\newcommand{\Lin}{\Link^{\text{in}}}
\newcommand{\kibitz}[2]{\ifnum\Comments=0\textcolor{#1}{#2}\fi}

\newcommand{\ubar}[1]{\underaccent{\bar}{#1}}

\newcommand{\real}{\mathbb{R}}

\newcommand{\mc}{\mathcal}

\newcommand{\lambdaup}{\bar{\lambda}}
\newcommand{\lambdalb}{\ubar{\lambda}}

\begin{document}
\maketitle

\begin{abstract}
	This paper designs traffic signal control policies for a network of signalized intersections without knowing the demand and parameters.
	Within a model predictive control (MPC) framework, control policies consist of an algorithm that estimates parameters and a one-step MPC that computes control inputs using estimated parameters.
	The algorithm switches between different terminal sets of the MPC to explore different regions of the state space, where different parameters are identifiable. 
	The one-step MPC minimizes a cost that approximates the sum of squares of all the queue lengths within a constant and does not require demand information.
	We show that the algorithm can estimate parameters exactly in finite time, and the one-step MPC renders maximum throughput in terms of input-to-state practical stability.
	Simulations indicate better transient performance regarding queue lengths under our proposed policies than existing ones.

\end{abstract}

\section{Introduction}
Signalized intersections in urban traffic networks account for a significant portion of congestion.
Consequently, effective traffic signal control is crucial for improving traffic flow and reducing travel time.
However, in practice, most existing traffic signals are fixed-time, often revised every few years, and hence need to be more adaptive to real-time traffic conditions.
With rapid advancements in traffic sensing through loop detectors and connected vehicles, 
there have been great interest and opportunities in developing and implementing dynamic traffic signal control policies.

There are two main approaches to dynamic traffic signal control: optimal control and queuing-theoretic methods.
Optimal control methods formulate the signal control as an optimization problem that minimizes a cost function, usually related to total queue lengths.
Dynamic programming, see \cite{cai2009adaptive}, and MPC, see \cite{aboudolas2010rolling}, have been used to solve the optimization problem recursively.
\cite{liu2023model} analyzed stabilizing properties of the MPC approach and showed exponential stability of a given equilibrium.
The controller requires complete knowledge of the demand and network parameters, and only small demand uncertainty is allowed for stability guarantees. 
In practice, external demand varies and the exact value is usually unknown to the controller.
In contrast to optimal control methods, queuing-theoretic methods, such as max-pressure or backpressure control, see \cite{varaiya2013max,gregoire2014capacity}, and proportional fair policy, see \cite{nilsson2015entropy,nilsson2020generalized,hosseini2016comparison},
do not require knowledge of demand. 
Additionally, the proportional fair policy also relaxes the requirement of knowing the network parameters.

In performance evaluation, one of the main metrics is throughput, which is measured in terms of the outflow from a network; it is equal to the inflow demand if queue lengths are bounded. The throughput \emph{region}, that is, all the possible throughput provided by a given signal controller, is characterized by the set of demands for which queue lengths are bounded under the given controller, starting from any initial condition. A controller is said to maximize throughput if the throughput region of the controller contains the throughput region of every other controller. It is shown in \cite{varaiya2013max} and \cite{nilsson2015entropy} that max-pressure and proportional fair policies maximize throughput, respectively. 
The key is due to the choice of cost functions that each policy optimizes. Specifically, max-pressure policy maximizes the difference between the upstream and downstream queue lengths, and proportional fair policy maximizes an entropy-like objective.
However, these cost functions are not directly related to economic objectives, such as delay or travel time, which are related to the norm of queue lengths.
If we change the cost functions, max-pressure and proportional fair policies are not guaranteed to maximize throughput anymore.
In general stochastic queuing networks, the drift-plus-penalty approach \cite{neely2006energy,neely2022stochastic}, was proposed to incorporate other performance metrics into the cost function, such as power usage in data networks, while guaranteeing maximum throughput.
However, the cost function needs to be independent of the queue lengths.

Our goal is to propose an MPC controller that minimizes the norm of queue lengths and maximizes throughput simultaneously with unknown demand and parameters.
We adopt deterministic queuing dynamics from \cite{varaiya2013max}, which is a discrete-time counterpart of the continuous-time model considered in \cite{nilsson2015entropy,nilsson2020generalized}.
The queuing dynamics are piecewise affine and more realistic than the linear dynamics considered in \cite{liu2023model}.
The MPC optimization is formulated with augmented state variables that serve as upper and lower bounds for the actual states.
A similar augmented idea was used in the authors' recent work on freeway traffic control, see \cite{li2024outputfeedbackadaptivemodelpredictive}, while the dynamics and objectives are different.

The contributions of this paper are as follows. 
First, given the upper and lower bounds of unknown demand and parameters,  i.e., saturation flow rates and turn ratios,
we propose an algorithm that is guaranteed to learn the parameters of the traffic network in finite time in feedback with the adaptive MPC controller.
By modifying the terminal sets in MPC, 
the algorithm explores and steers the system to different regions of the state space in which corresponding parameters can be estimated exactly.
Our second contribution is a one-step MPC optimization problem that only uses parameters learned from the algorithm and does not require the knowledge of unknown demand.
One-step MPC for traffic signal timing has been used in \cite{grandinetti2018distributed} where the focus is on convex relaxation and demand knowledge is still required. 
Optimal solutions to the one-step MPC are shown to minimize a cost function that approximates the 2-norm of the queue lengths within a constant. 
In addition, the one-step MPC can be implemented in a distrusted manner similar to the max-pressure policy.
Third, we show that the one-step MPC is maximally stabilizing, i.e., the closed-loop system in feedback with the one-step MPC is input-to-state practically stable under exogenous demand within a set, which is the largest where stability is possible.
This set is characterized following \cite{varaiya2013max}.

We conclude this section by defining the notation used throughout the paper.
Let $\real,\real_{\geq0},\real_{>0}$ denote the sets of real, nonnegative real, and positive real numbers, respectively.
For a finite set $\mc A$, let $|\mc A|$ denote the cardinality of $\mc A$, 
and let $\mc S_{\mc A}:=\{x\in\real_{\geq0}^{|\mc A|}:\sum_{i\in\mc A}x_i=1\}$ denote the simplex over the elements of $\mc A$.
A continuous function $\alpha: \real_{\geq0}\to \real_{\geq0}$ is a $\mc K$-function if $\alpha(0) = 0,\alpha(s) > 0$
for all $s > 0$, and it is strictly increasing. A function
$\alpha(\cdot)$ is a $\mc K_{\infty}$-function if it is a $\mc K$-function and $\alpha(s) \to\infty$ when $s\to\infty$. 
A continuous function $\beta:\real_{\geq0}\times \mathbb{Z}_{\geq0} \to\real_{\geq0}$ is a
$\mc {KL}$-function if $\beta(s,t)$ is a $\mc K$-function in $s$ for every $t\geq0$, 
it is strictly decreasing in $t$ for every $s > 0$, and $\beta(s, t)\to 0$ when $t \to\infty$.
\section{Problem Formulation}
\begin{figure}[tb!]
	\centering
	\includegraphics[width=0.44\textwidth]{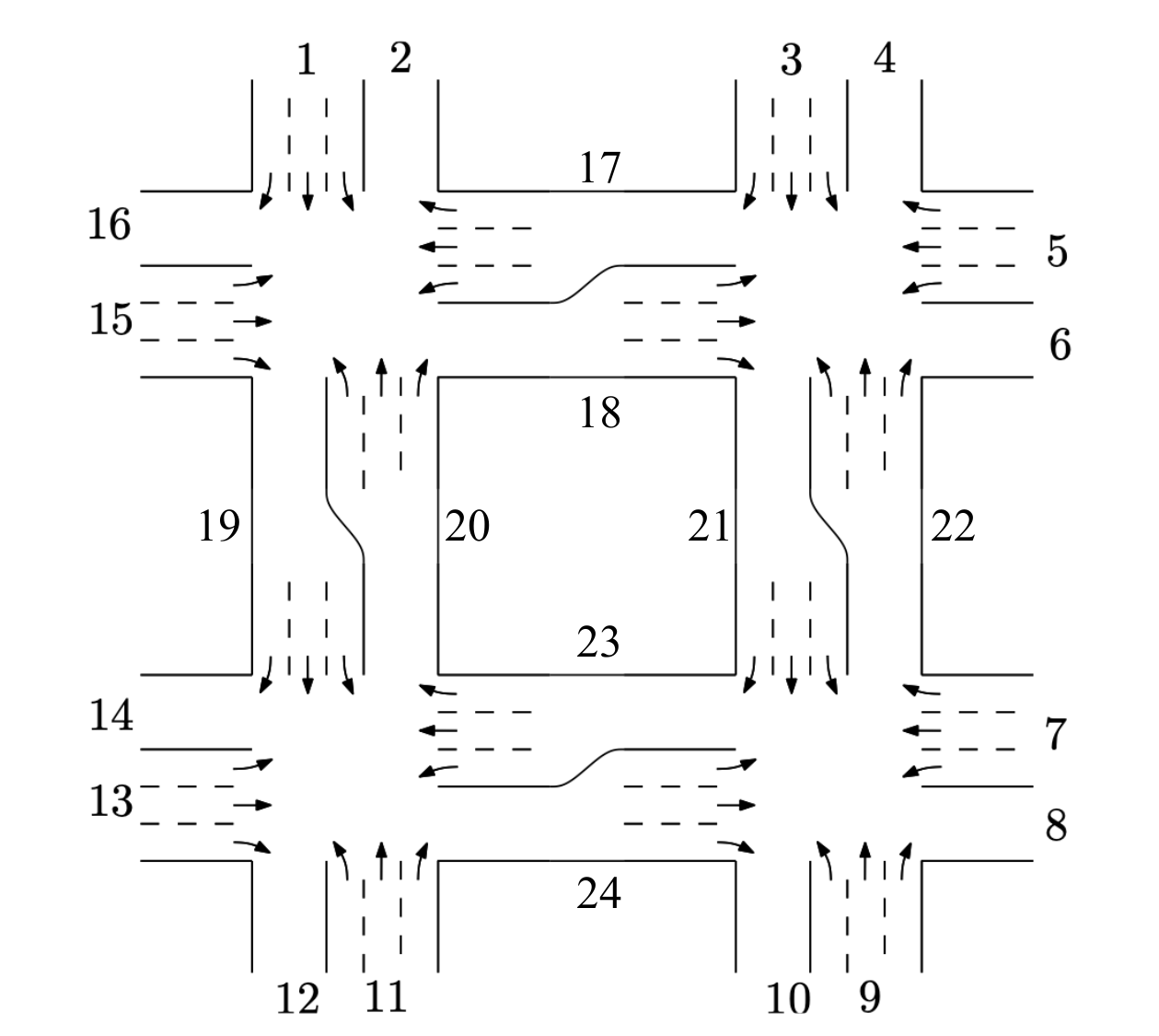}
	\caption{A $4\times4$ grid network}
	\label{fig:traffic-network}
\end{figure}
Let $\mathcal{G}=(\mathcal{N}, \Link)$ denote a directed traffic flow network where $\mathcal{N},\Link$ denote the finite set of nodes and links, respectively.
Let $\Lint$ denote the set of internal links that connects two nodes in $\mathcal{N}$; 
let $\Lentry$ denote the set of entry links that have no start node in $\mathcal{N}$;
let $\Lexit$ denote the set of exit links that have no end node in $\mathcal{N}$.
Fix a link $i\in\Link$, let $\Lin_i$ denote the set of links input to $i$, and let $\Lout_i$ denote the set of links output from $i$; 
for all $j\in\Link^{\text{out}}_i$, let $x_{ij}\in\real_{\geq0}$ denote the length of the queue that goes from link $i$ to link $j$; let $S_{ij}(t)\in[0,1]$ denote the fraction of green time at time $t$ for the movement from $i$ to $j$; 
let $R_{ij}(t)\in[0,1]$ denote the fraction of the queue at link $i$ that goes to link $j$;
let $C_{ij}\in\real_{\geq0}$ denote the rate of saturation flow from $i$ to $j$.
For all $i\in\Lentry$, let $\lambda_{i}(t)\in\real_{\geq0}$ denote the rate of exogenous demand entering link $i$.
We consider the traffic dynamics from \cite{varaiya2013max}:
\begin{equation}\label{eq:dynamics}
	\begin{aligned}
		&x_{ij}(t+1) = \max\{x_{ij}(t) - C_{ij}S_{ij}(t),0\} \\  & \hspace{0.1in} + \sum_{k\in\Lin_i}R_{ij}\min\{C_{ki}S_{ki}(t), x_{ki}(t)\}, i\in\Link_{\text{int}},j\in\Link^{\text{out}}_i\\
		&x_{ij}(t+1) = \max\{x_{ij}(t) - C_{ij}S_{ij}(t),0\}  + R_{ij}\lambda_{i}(t),\\ & \hspace{2in} i\in\Link_{\text{entry}}, j\in\Link^{\text{out}}_i \\
		&x_{i}(t+1) = \sum_{k\in\Lin_i}\min\{C_{ki}S_{ki}(t), x_{ki}(t)\}, i \in\Link_{\text{exit}} 
	\end{aligned}
\end{equation}

\begin{example}
	For the $4\times4$ grid network shown in Figure~\ref{fig:traffic-network}, $\Link = \{1,2,\ldots,24\}$, $\Lint = \{17,18,\ldots,24\}$, $\Lentry = \{1,3,5,7,9,11,13,15\}$, $\Lexit = \{2,4,6,8,10,12,14,16\}$.
\end{example}

Let $x(t) := \{x_{ij}(t), i\in\Lint\cup\Lentry, j\in\Lout_i \}\in\real^{n_x}_{\geq0}$ denote the vector of queue lengths excluding the exit links, and $\lambda(t):=\{\lambda_i(t),  i\in\Lentry\}$ denote the vector of demand.
The queue lengths of exit links can be computed from the queue lengths of entry and internal links.
Let $S_{ij}(t)=0, C_{ij}=0, R_{ij}=0$ if $i$ and $j$ are not connected.
Let $S(t), C, R$ be the compact notations for the matrices of  signal control, saturation flow rates, and turn ratios, respectively.
\begin{figure}
	\centering
	\includegraphics[width=0.48\textwidth]{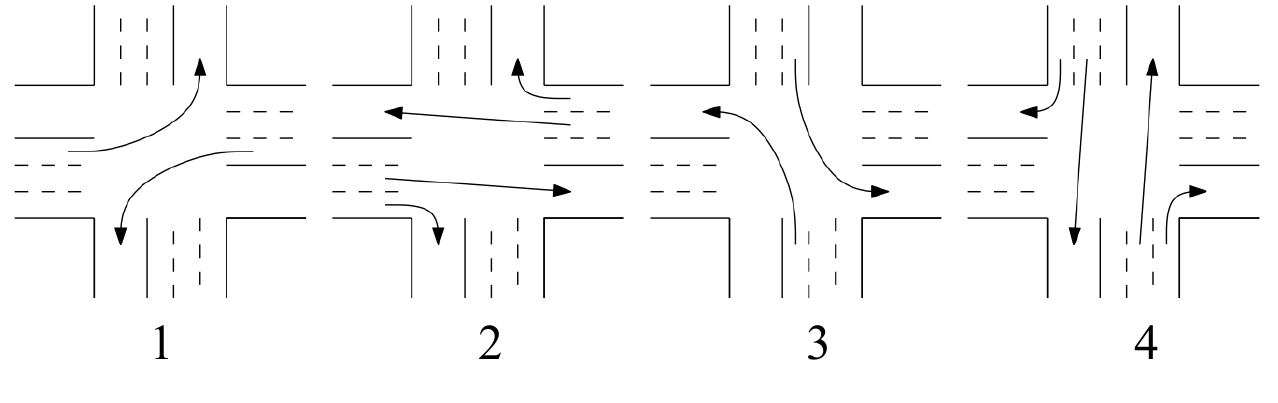}
	\caption{Illustration of a four-phase architecture at an
	intersection}
	\label{fig:phases}
\end{figure}
By definition, the routing matrix $R$ satisfies (i) $R_{ij}=0$ for all $i\in\Link, j\notin\Lout_i$, (ii) $\sum_{j\in\Link}R_{ij}=1$ for all $i\in\Lint\cup\Lentry$.
According to \cite[Proposition 1]{varaiya2013max}, for every constant demand vector $\lambda$, there is a unique flow $q:=\{q_i,i\in\Link\}$ that satisfies:
\begin{equation}\label{eq:flow}
	\begin{aligned}
		q_i = & \lambda_i, && i\in\Lentry \\
		q_i = & \sum_{j\in\Link}R_{ij}q_j, && i\in\Lint\cup\Lexit
	\end{aligned}
\end{equation}

We assume the signal control for each intersection activates in phases. 
For each intersection $n\in\mathcal{N}$, a phase is a set of queues $(i,j)$ that are connected through $n$ and allowed to move simultaneously. 
Let $\mc P(n)$ be the set of phases at node $n$.
An example of a four-way intersection with four phases can be found in Figure~\ref{fig:phases}.
Let $n_u$ be the total number of phases for all intersections.
Let $u(t)\in\mathcal{U}\subset\real_{\geq0}^{n_u}$ denote the vector of split ratios for all phases at time $t$. 
The set $\mathcal{U}$ is the union of the sets\/ $\mc S_{\mc P(n)}$ for all $n\in\mathcal{N}$.
Let $S^m$ denote the signal control matrix for phase $m$; $S^m_{ij} = 1$ if queue $(i,j)$ is served during phase $m$ and $S^m_{ij}=0$ otherwise.
We make the following assumption on the signal control matrices:
\begin{assumption}\label{assumption:signal}
	For all $i,j=1,\ldots,|\Link|$:
	\begin{enumerate}
		\item $\sum_{m=1}^{n_u}S^m_{ij}\geq1$ if $i$ and $j$ are connected.
		\item Let $n_j$ be the start node of link $j$. Then, there exists a phase $m\in\mc P(n_j)$ such that $\sum_{i\in\Link}S^m_{ij}=0$.
	\end{enumerate}
\end{assumption}
\begin{remark}
	The first condition in Assumption~\ref{assumption:signal} ensures that each queue is activated at least once during all phases.
	The second condition ensures that for each downstream link at every intersection, there is at least one phase that does not serve any queue entering that downstream link.
	This is common in practice to avoid conflicts between movements.
\end{remark}

The signal control matrices $S(t)$ is defined as $S(t) \equiv S(u(t))= \sum_{m=1}^M u_m(t)S^m$.
The dynamics \eqref{eq:dynamics} can then be compactly written as 
$$x(t+1) = f(x(t),u(t),\lambda(t);C,R)$$

Since the control set $\mathcal{U}$ is bounded, the unique flow defined in \eqref{eq:flow} may not be achieved if the demand is large. 
We consider the following set of demand such that the unique flow $q$ is \textit{feasible}:
\begin{equation}\label{eq:feasible-demand}
	\begin{aligned}
		\Lambda := \{\lambda\in\real_{\geq0}^{|\Lentry|}:\ &\exists \ u\in\mathcal{U} \text{ s.t. } C_{ij}S_{ij}(u)>q_iR_{ij},\\
		&\forall\, i\in\Lint\cup\Lentry, j\in\Lout_i\}
	\end{aligned}
\end{equation}

Our goal is to design a controller that stabilizes every queue of the traffic network starting from any initial condition in $\real_{\geq0}^{n_x}$ for all $\lambda\in\Lambda$, and thus renders maximal throughput. 
Existing control policies such as max-pressure control \cite{varaiya2013max} or proportional fair policy are able to achieve this goal. 
However, these two policies do not take other performance into account in addition to throughput, such as total travel time.
Instead, we follow the model predictive control (MPC) scheme, which allows other performance metrics to be directly incorporated as cost functions.
Given forward horizon $T$, running cost $\ell:\real^{n_x}\times\real^{n_u}\to\real_{\geq0}$, and terminal cost $V_f:\real^{n_x}\to\real_{\geq0}$, the MPC approach solves the following finite-horizon optimal control problem:
\begin{equation}\label{eq:mpc-opt}
	\min_{u(0|t), \ldots, u(T-1|t)}\sum_{\tau=0}^{T-1} \ell(x(\tau|t),u(\tau|t)) + V_f(x(T|t))
\end{equation}
subject to dynamics \eqref{eq:dynamics} with initial condition $x(0|t) = x(t)$, $u(\tau|t)\in\mathcal{U}, \tau=0,\ldots,T-1$, and terminal constraint $x(T|t)\in\mc X_f$.
Let $\{\hat{u}^*(0|t),\ldots,\hat{u}^*(T-1|t)$ denote an optimal solution to \eqref{eq:mpc-opt},
 we set $u(t)$ to be equal to $\hat{u}^*(0|T)$, then \eqref{eq:mpc-opt} is solved at $t+1$ to similarly obtain $u(t+1)$, and so on.

The optimization \eqref{eq:mpc-opt} assumes that the dynamics \eqref{eq:dynamics} are known, i.e., saturation ratio $C$, routing matrix $R$, and demand $\lambda(t)$ are available to the controller.
Since the exact knowledge of the parameters $C, R$ and exogenous demand $\lambda(t)$ are usually unknown,
we design an algorithm that learns parameters $R$ and $C$ in feedback with the MPC controller \eqref{eq:mpc-opt}. 
We also relax the requirement of knowing demand $\lambda(t)$ in the controller.

\section{Adaptive MPC}

We first introduce auxiliary variables for parameters $C, R$ and demand $\lambda$. 
For all $i\in\Lint\cup\Lentry, j\in\Lout_i$, let $\bar{C}_{ij}$ and $\ubar{C}_{ij}$ denote an upper and lower bound for $C_{ij}$, respectively;
let $\bar{R}_{ij}$ and $\ubar{R}_{ij}$ denote an upper and lower bound for $R_{ij}$, respectively.
For all $i\in\Lentry$, let $\bar{\lambda}_i$ and $\ubar{\lambda}_i$ denote an upper and lower bound for $\lambda_i(t)$, respectively.
The relations between the auxiliary variables and the actual parameters are summarized as follows:
\begin{assumption}\label{assumption:bounds}
	For all $i,j=1,\ldots,|\Link|$, we assume 
	\begin{enumerate}
	\item $0<\ubar{C}_{ij}\leq C_{ij}\leq\bar{C}_{ij}$ if $C_{ij}>0$ and $\ubar{C}_{ij} = \bar{C}_{ij}=0$ if $C_{ij} = 0$;
	\item $0<\ubar{R}_{ij}\leq R_{ij}\leq\bar{R}_{ij}$ if $R_{ij}>0$ and $\ubar{R}_{ij} = \bar{R}_{ij}=0$ if $R_{ij} = 0$;
	\item $0\leq\lambdalb_i\leq\lambda_i(t)\leq\lambdaup_i$ for all $t\geq0$ if $i\in\Lentry$;
	\item $\bar{R}_{ij}\bar{\lambda}_i < \ubar{C}_{ij}$ for all $i\in\Lentry, j\in\Lout_i$.
	\end{enumerate}
	Additionally, the upper and lower bounds are known to the controller for all $t\geq0$.
\end{assumption}
\begin{remark}
	The upper and lower bounds required in Assumption~\ref{assumption:bounds} can be obtained from historical data.
	The fourth condition is not too restrictive in practice since maximal demand is usually less than the saturation flow rate of entry links.
\end{remark}

With the auxiliary variables, we consider the following augmented dynamics:
\begin{equation}\label{eq:augmented-dynamics}
	\begin{aligned}
		\bar{x}(t+1) &= F(\bar{x}(t), u(t), \lambdaup;\bar{C},\ubar{C},\bar{R}), \\
		\ubar{x}(t+1) &= F(\ubar{x}(t), u(t), \lambdalb;\ubar{C},\bar{C},\ubar{R})
	\end{aligned}
\end{equation}
where 
\begin{equation*}
	\begin{split}
		&F_{ij}(x,u,\lambda;C,\tilde{C},R) :=  \max\{x_{ij} - \tilde{C}_{ij}S_{ij}(u),0\} \\ & \hspace{0.3in}+ \sum_{k\in\Link}R_{ij}\min\{C_{ki}S_{ki}(u), x_{ki}\}, \quad i\in\Link_{\text{int}},j\in\Link^{\text{out}}_i\\
		&F_{ij}(x,u,\lambda;C,\tilde{C},R) :=  \max\{x_{ij} - \tilde{C}_{ij}S_{ij}(u),0\}  + R_{ij}\lambda_{i}, \\ &\hspace{2.3in} i\in\Link_{\text{entry}}, j\in\Link^{\text{out}}_i \\
	\end{split}
\end{equation*}
The only difference between augmented dynamics $F$ in \eqref{eq:augmented-dynamics} and the original dynamics $f$ in \eqref{eq:dynamics} is that the saturation flow rate $C$ with a negative sign is replaced by the auxiliary variable $\tilde{C}$.
With Assumption~\ref{assumption:bounds}, the state trajectories generated by the augmented dynamics $F$ are guaranteed to be upper and lower bounds for the actual states, i.e., $\ubar{x}(t)\leq x(t)\leq\bar{x}(t)$ for all $t\geq0$ if $\ubar{x}(0)\leq x(0)\leq \bar{x}(0)$.
Then, we propose the following augmented MPC problem:
\begin{equation}\label{eq:mpc-augmented}
	\begin{aligned}
		&\min_{u(0|t), \ldots, u(T-1|t)}\sum_{\tau=0}^{T-1} \ell(\bar{x}(\tau|t),u(\tau|t))  \\
		\text{s.t. } & \bar{x}(\tau+1|t) = F(\bar{x}(\tau|t),u(\tau|t),\lambdaup;\bar{C},\ubar{C},\bar{R}), \\ & \hspace{2in}\tau=0,\ldots,T-1 \\
		& \ubar{x}(\tau+1|t) = F(\ubar{x}(\tau|t),u(\tau|t),\lambdalb;\ubar{C},\bar{C},\ubar{R}), \\ & \hspace{2in} \tau=0,\ldots,T-1 \\
		& \bar{x}(0|t) = x(t), \ubar{x}(0|t) = x(t), (\bar{x}(T|t),\ubar{x}(T|t))\in\mc X_f \\
		& u(\tau|t) \in \mc U, \quad \tau=0,\ldots,T-1
	\end{aligned}
\end{equation}
Let $\mc X_f=\{\bar{x},\ubar{x}\in\real_{\geq0}^{n_x}:\exists\, u\in\mc U \text{ s.t. }(\bar{x},\ubar{x})\in\mc X_u(u)\}$ where $\mc X_u(u)$ denotes a set of states that depends on $u$.
The set $\mc X_u(u)$ will be constructed such that nonzero entries in $C$ or $R$ can be uniquely determined.
Based on the augmented MPC \eqref{eq:mpc-augmented}, we propose Algorithm~\ref{alg:adpative-mpc} that iteratively solves \eqref{eq:mpc-augmented} with different terminal constraints, see Section~\ref{sec:algorithm}.
Algorithm~\ref{alg:adpative-mpc} consists of a function $\textbf{MPC}(\mc X_u(u))$ and a loop over all pairs $(i,j)$. 
For each pair $(i,j)$ in the loop, Algorithm~\ref{alg:adpative-mpc} determines a set $\mc X_u(u)$ such that either $R_{ij}$ or $C_{ij}$ can be uniquely determined when $x\in\mc X_u(u)$ for some $u\in\mc U$. 
Given $\mc X_u(u)$, the function $\textbf{MPC}(\mc X_u(u))$ constructs a terminal set and runs dynamics \eqref{eq:dynamics} in feedback with \eqref{eq:mpc-augmented} until $x(t)\in\mc X_u(u)$ for some $u\in\mc U$.
We will show that by repeating the process, parameters $C$ and $R$ can be exactly estimated in finite time.
\begin{theorem}
	Let Assumptions~\ref{assumption:signal} and \ref{assumption:bounds} hold. 
	Suppose $\ell(\bar{x},u)=0$ if $(\bar{x},\ubar{x})\in\mc X_f$ and $\ell(\bar{x},u) = l^T\bar{x}$ otherwise with cost coefficient $l\in\real^{n_x}_{>0}$.
	Then, Algorithm~\ref{alg:adpative-mpc} terminates in finite time and the outputs satisfy $\ubar{C}_{ij} = C_{ij} = \bar{C}_{ij}$ for all $i\in\Lint\cup\Lentry, j\in\Lout_i$, and $\ubar{R}_{ij} = R_{ij} = \bar{R}_{ij}$ for all $i\in\Lint, j\in\Lout_i$.
\end{theorem}
\begin{remark}
	The augmented MPC \eqref{eq:mpc-augmented} can be reformulated as a mixed-integer linear program since the dynamics $F$ are piecewise affine and the terminal constraints considered in Algorithm~\ref{alg:adpative-mpc} are linear.
\end{remark}
 
After Algorithm~\ref{alg:adpative-mpc} terminates, saturation flow rates for all links and turn ratios for internal links are exactly estimated while demand values are still unknown. 
To avoid the use of unknown demand, we consider the following one-step MPC optimization:
\begin{equation}\label{eq:mpc-optimization}
	\begin{aligned}
		&\min_{u(0|t)\in\mc U}
		\sum_{i\in\Link_{\text{entry}}}\sum_{j\in\Link^{\text{out}}_i} C_{ij}^2S_{ij}^2(u(0|t)) - 2C_{ij}S_{ij}(u(0|t))x_{ij}(t)\\
		 &\qquad \quad+  \sum_{i\in\Link_{\text{int}}}\sum_{j\in\Link^{\text{out}}_i}x_{ij}^2(1|t) \\
		 &\text{s.t.}\ x_{ij}(1|t) = \max\{x_{ij}(t) - C_{ij}S_{ij}(t),0\} \\  & \hspace{0.35in} + \sum_{k\in\Lin_i}R_{ij}\min\{C_{ki}S_{ki}(t), x_{ki}(t)\}, i\in\Link_{\text{int}},j\in\Link^{\text{out}}_i\\
		\end{aligned}
\end{equation}
\begin{remark}\label{remark:cost-queue-length-relations}
	Let $\epsilon>0$ be a constant.
	Solving \eqref{eq:mpc-optimization} is the equivalent of solving the following
	\begin{equation}\label{eq:mpc-optimization-lyapunov}
		\begin{aligned}
			&\min_{u(0|t)\in\mc U}
			\epsilon \|x(t)\|_1 + \|\lambda\|_2^2 \\
			&+\sum_{i\in\Link_{\text{entry}}}\sum_{j\in\Link^{\text{out}}_i}
			x^2_{ij}(t) + 2R_{ij}\lambda_ix_{ij}(t)\\ 
			&+\sum_{i\in\Link_{\text{entry}}}\sum_{j\in\Link^{\text{out}}_i}
			 C_{ij}^2S_{ij}^2(u(0|t)) - 2C_{ij}S_{ij}(u(0|t))x_{ij}(t)\\
			 &+  \sum_{i\in\Link_{\text{int}}}\sum_{j\in\Link^{\text{out}}_i}x_{ij}^2(1|t) \\
			 \text{s.t.}&\quad x_{ij}(1|t) = f_{ij}(x(t),u(0|t),\lambda;C,R),i\in\Link_{\text{int}},j\in\Link^{\text{out}}_i\\
		\end{aligned}
	\end{equation}
	since only constant parameters are added to the objective function, and thus the sets of optimal solutions are the same.
	The cost function of \eqref{eq:mpc-optimization-lyapunov} is also shown to approximate $\|x(1|t)\|_2^2$ within a constant in the proof of Theorem~\ref{theorem:stability}.
\end{remark}
\begin{remark}
	The optimization \eqref{eq:mpc-optimization} can also be solved in a distributed manner since the cost function and the constraints are separable with respect to each node.
	The controller at each node only needs the information from itself and its neighboring nodes.
\end{remark}

We will show that queue lengths of the closed-loop system \eqref{eq:dynamics} in feedback with MPC controller \eqref{eq:mpc-optimization} are bounded in terms of \emph{input-to-state practically stable (ISpS)}, see Section~\ref{sec:ISpS}.
\begin{theorem}\label{theorem:stability}
	Let $\lambda(t)\equiv \lambda\in\Lambda$ defined in \eqref{eq:feasible-demand}. 
	Then, the closed-loop system \eqref{eq:dynamics} in feedback with MPC controller \eqref{eq:mpc-optimization} is \emph{input-to-state practically stable} with respect to $\lambda$, i.e.,
	there exists a $\mc {KL}$ function $\beta(\cdot,\cdot)$, a pair of $\mc K$ functions $\delta_1(\cdot),\delta_2(\cdot)$, and a constant $\mu\geq\|\lambda\|$ for all $\lambda\in\Lambda$
	such that
	\begin{equation}
		\begin{aligned}
			&\|x(t)\| \leq \beta(\|x(0)\|,t) + \delta_1(\|\lambda\|_{\infty}) + \delta_2(\mu), \\ & \hspace{2in} \forall\, x(0)\in\real^{n_x}_{\geq0}, \quad t\geq0
		\end{aligned}
	\end{equation}
\end{theorem}
\begin{remark}
    Theorem~\ref{theorem:stability} implicitly assumes that parameters $C$ and $R$ are known, which can be obtained from Algorithm~\ref{alg:adpative-mpc}. Also, we do not have any requirements on phases as the feasible demand is specified by $\Lambda$ that depends on the given phase architecture. This is consistent with the max-pressure control, see \cite{varaiya2013max}.
\end{remark}

\section{Numerical simulations}
In this section, we compare the performance in terms of queue length through simulations between our proposed adaptive MPC, proportional fair, and max pressure policies.

We conduct numerical simulations on a $4\times4$ grid network with 24 links, as shown in Figure~\ref{fig:traffic-network}.
Every intersection uses the same phase architecture as shown in Figure~\ref{fig:phases}.
We follow the network parameters used in \cite{nilsson2015entropy} for which the .
For internal links, the turn ratios are assumed to be 0.17 for left turning, 0.33 for through movement, and 0.5 for right turning. 
For entry links, the turn ratios are assumed to be 1/3 for all movements.
The saturation flow rates are symmetrical throughout the network and are specified to be 1.5 for the left lane, 1.6 for the middle lane, and 1.7 for the right lane.
External demand $\lambda_i$ is set to be 0.93 for all $i\in\Lentry$. 
The initial queue length is set to be $x_{ij}(0)=1$ for all $i\in\Lint\cup\Lentry,j\in\Lout_i$. 
The optimization problems \eqref{eq:mpc-augmented} and \eqref{eq:mpc-optimization} are solved using the \texttt{Gurobi} solver.
\begin{figure}[tb]
	\centering
	\includegraphics[width=0.52\textwidth]{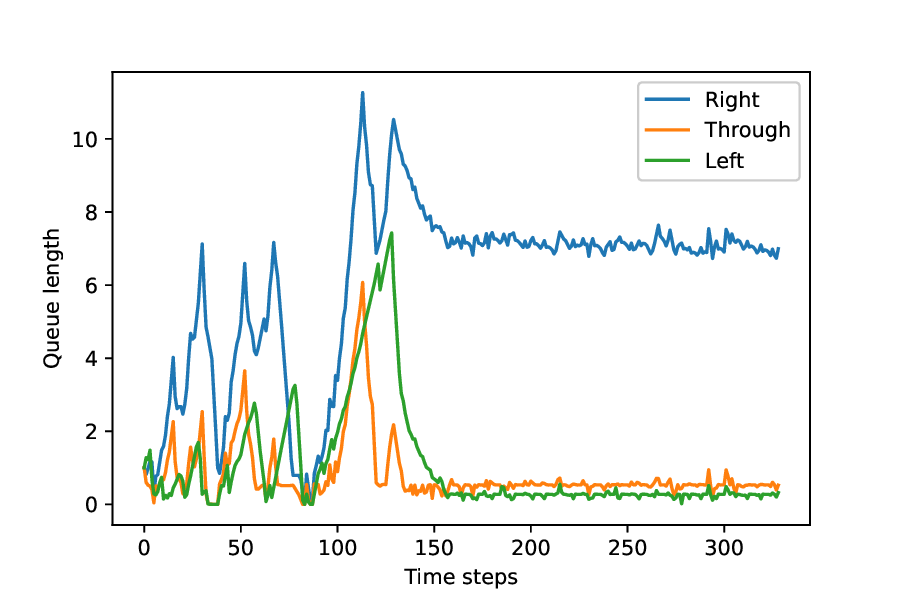}
	\caption{Evolution of queue lengths at link 20 under the adaptive MPC}
	\label{fig:termination}
\end{figure}

We first verify that Algorithm~\ref{alg:adpative-mpc} terminates in finite time.
We use the upper bounds $\bar{C}_{ij} = C_{ij} + 0.1, \bar{R}_{ij} = R_{ij} + 0.1$ 
and lower bounds $\ubar{C}_{ij} = C_{ij} - 0.1, \ubar{R}_{ij} = R_{ij} - 0.1$ for all $i\in\Lint\cup\Lentry, j\in\Lout_i$.	
For demand, we set $\bar{\lambda}_i = \lambda_i + 0.1, \ubar{\lambda}_i = \lambda_i - 0.1$ for all $i\in\Lentry$.
We run Algorithm~\ref{alg:adpative-mpc} with the given upper and lower bounds and use the one-step MPC \eqref{eq:mpc-optimization} after Algorithm~\ref{alg:adpative-mpc} terminates.
Figure~\ref{fig:termination} shows the evolution of queue lengths at link 20 which has large queue length for better illustration. 
All links exhibit similar trends.
It can be seen that Algorithm~\ref{alg:adpative-mpc} terminates between 100 and 150 steps. 
During Algorithm~\ref{alg:adpative-mpc}, the system trajectories do not follow a specific pattern since the controller is exploring the system to learn the parameters.
After the parameters are learned, the one-step MPC stabilizes the system and the queue lengths remain bounded. 
\begin{figure}[tb]
	\centering
	\includegraphics[width=0.52\textwidth]{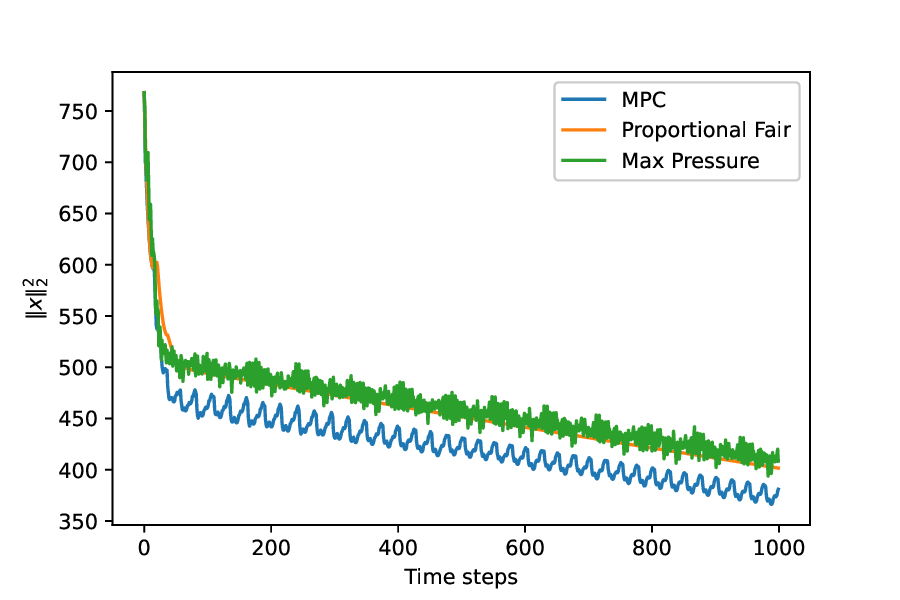}
	\caption{Comparison of $\|x\|_2^2$ under different control policies}
	\label{fig:comparison}
\end{figure}

Then, we show that the one-step MPC \eqref{eq:mpc-optimization} results in smaller norm of queue lengths compared to other queuing-theoretic policies due to the choice of cost function (c.f. Remark~\ref{remark:cost-queue-length-relations}). 
Figure~\ref{fig:comparison} shows the evolution of $\|x\|_2^2$, starting from the same initial conditions under one-step MPC in \eqref{eq:mpc-optimization}, max-pressure, and proportional fair policies.
The MPC and max-pressure controllers are implemented given parameters $C$ and $R$.
It can be seen that our MPC controller results in the smallest 2-norm of queue lengths among all policies in the transient phase.

\section{Conclusions}

In this paper, we design an MPC-based framework for traffic signal control. 
The framework consists of an adaptive MPC algorithm that iteratively learns the parameters using various terminal sets and a one-step MPC optimization that approximately minimizes queue lengths without using demand.
We prove that this framework is guaranteed to learn the parameters exactly in finite time and maximally stabilize the system in terms of input-to-state practical stability.
The numerical simulations suggest that our proposed MPC controller outperforms other queuing theoretic policies in terms of queue length.
Future work includes extending the proposed framework to be responsive to parameter changes and output-feedback settings when the queue lengths are not directly measured. The one-step MPC can be reformulated as a convex optimization problem to reduce the computational complexity following the approach in \cite{grandinetti2018distributed}.
Also, it is interesting to extend the MPC approach in stochastic queuing networks.

\bibliographystyle{ieeetr}
\bibliography{references.bib}

\section*{Appendix}
\subsection{Algorithm}
\label{sec:algorithm}
\begin{algorithm}
	\caption{Adaptive MPC}
	\label{alg:adpative-mpc}
	\begin{algorithmic}[1]
		\STATE \textbf{Input:} Upper and lower bounds of demand and parameters specified in Assumption~\ref{assumption:bounds}.
		\STATE \textbf{Output:} $\bar{C}_{ij},\ubar{C}_{ij}$ for all $i\in\Lint\cup\Lentry,j\in\Lout_i; \bar{R}_{ij},\ubar{R}_{ij}$ for all $i\in\Lint,j\in\Lout_i$.  
	\end{algorithmic}
\end{algorithm}
\begin{algorithm}
	\begin{algorithmic}[1]
		\setcounter{ALC@line}{2}
		\STATE \textbf{function MPC}$(\mc X_u(u))$	
		\begin{ALC@g}
			\STATE Set $\mc X_f=\{\bar{x},\ubar{x}\in\real_{\geq0}^{n_x}:\exists\, u\in\mc U \text{ s.t. }(\bar{x},\ubar{x})\in\mc X_u(u)\}$
			\STATE Choose $T$ such that \eqref{eq:mpc-augmented} is feasible
			\WHILE{$x(t)\notin\mc X_f$}
				\STATE Solve the MPC optimization \eqref{eq:mpc-augmented} to obtain an optimal solution $\{\hat{u}^*(0|t),\ldots,\hat{u}^*(T-1|t)\}$.
				\STATE Set $u(t) = \hat{u}^*(0|t)$ and update $x(t+1)$
				\STATE Set $t\leftarrow t+1$
			\ENDWHILE
			\STATE Find $u\in\mc U$ such that $x(t)\in\mc X_u(u)$
			\STATE Set $u(t) = u$ and update $x(t+1)$.
			\STATE Set $t\leftarrow t+1$.
		\end{ALC@g}
		\STATE \textbf{end function}
		\STATE Initialize $t=0$, $x(0)$, $\bar{C}$, $\ubar{C}$, $\bar{R}$, $\ubar{R}$, $\bar{\lambda}$, $\ubar{\lambda}$.
		\FOR{$i\in\Lint\cup\Lentry$}
			\FOR{$j\in\Lout_i$}
				\IF{$\bar{R}_{ij} \neq \ubar{R}_{ij}, i\in\Lint$}
					\STATE Set $\mc X_u(u) = \{\bar{x},\ubar{x}\in\real_{\geq0}^{n_x}\colon\bar{x}_{ij}\leq \ubar{C}_{ij}S_{ij}(u), \bar{x}_{ki}\leq \ubar{C}_{ki}S_{ki}(u)\ \forall\, k\in\mc L^{\text{in}}_i\}$
					\STATE Run $\textbf{MPC}(\mc X_u(u))$
					\STATE Set $\bar{R}_{ij} = \ubar{R}_{ij} = \dfrac{\sum_{k\in\Link}x_{ki}(t-1)}{x_{ij}(t)}$.
				\ELSIF{$\bar{C}_{ij} \neq \ubar{C}_{ij}, i\in\Lint$}
					\STATE Set $\mc X_u(u) = \{\bar{x},\ubar{x}\in\real_{\geq0}^{n_x}\colon\ubar{x}_{ij}\geq \bar{C}_{ij}S_{ij}(u), \bar{x}_{ki}\leq \ubar{C}_{ki}S_{ki}(u)\ \forall\, k\in\mc L^{\text{in}}_i\}$
					\STATE Run $\textbf{MPC}(\mc X_u(u))$
					\STATE Set $\bar{C}_{ij} = \ubar{C}_{ij} = \dfrac{x_{ij}(t) - x_{ij}(t - 1) - \sum_{k\in\Link}R_{ij}x_{ki}(t-1)}{S_{ij}(u(t-1))}$.
				\ENDIF
				\IF{$\bar{C}_{ij}\neq \ubar{C}_{ij}, i\in\Lentry, j\in\Lexit$}
					\STATE Set $\mc X_u(u) = \{\bar{x},\ubar{x}\in\real_{\geq0}^{n_x}\colon \ubar{x}_{ij}\geq\bar{C}_{ij}S_{ij}(u),\ \bar{x}_{kj} \leq C_{kj}S_{kj}(u), \forall\, k\in\Lin_j\setminus\{i\} \}$.
					\STATE Run $\textbf{MPC}(\mc X_u(u))$
					\STATE Set $\bar{C}_{ij} = \ubar{C}_{ij} = \dfrac{x_j(t)-\sum_{k\in\Lin_j\setminus\{i\}}x_{kj}(t-1)}{S_{ij}(u(t-1))}$. 
				\ELSIF{$\bar{C}_{ij}\neq \ubar{C}_{ij}, i\in\Lentry, j\in\Lint$}
					\STATE Choose $l$ such that $\ubar{C}_{jl} = \bar{C}_{jl}$. 
					\STATE Set $\mc X_u(u) = \{\bar{x},\ubar{x}\in\real_{\geq0}^{n_x}\colon \ubar{x}_{ij}\geq\bar{C}_{ij}S_{ij}(u),\ \bar{x}_{kj} \leq \ubar{C}_{kj}S_{kj}(u), \forall\, k\in\Lin_j\setminus \{i\}\}$.
					\STATE Run $\textbf{MPC}(\mc X_u(u))$
					\STATE Set $\bar{C}_{ij} = \ubar{C}_{ij} = \big[x_{jl}(t)-\max\{x_{jl}(t-1)-\bar{C}_{jl}S_{jl}(u(t-1)),0\}-\sum_{k\in\Lin_j\setminus\{i\}}x_{kj}(t-1)\big]/S_{ij}(u(t-1))$.
				\ENDIF
			\ENDFOR
		\ENDFOR
	\end{algorithmic}
\end{algorithm}
\subsection{Max-pressure control}
Let $u^{\text{mp}}(x)$ denote the max-pressure control proposed in \cite{varaiya2013max}. 
The max-pressure controller satisfies the following property:
	\begin{lemma}\label{lemma:max-pressure}
		(\cite[Theorem 2]{varaiya2013max})
		Consider the traffic dynamics \eqref{eq:dynamics}. Let $\lambda\in\Lambda$. Then, there exists $k<\infty$ and $\epsilon>0$ such that
		\begin{equation}
			\|f(x,u^{\text{mp}}(x),\lambda;R,C)\|_2^2 - \|x\|_2^2 \leq -\epsilon\|x\|_1 + k, \quad \forall\, x\in\real^{n_x}_{\geq0}
		\end{equation}
	\end{lemma}
\subsection{Input-to-state practical stability}
\label{sec:ISpS}
Consider a nonlinear dynamical system of the form 
\begin{equation}\label{eq:appendix-dynamics}
	x^+ = g(x,w)
\end{equation}
where $x\in\real^n$ is the system state, $w\in\real^q$ is the disturbance input, and $x^+$ is the successor state.
The disturbance input $w$ satisfies
\begin{equation}\label{eq:appendix-disturbance}
	w\in\mc W = \{w\in\real^q: \|w\|\leq \rho(\mu)\}
\end{equation}
where $\mu\geq0$ is a constant, and $\rho(\cdot)$ is a $\mc K$-function.
Now we state the input-to-state practical stability (ISpS) results from \cite{limon2006input}. Let $\|\cdot\|$ be any $p$-norm.
\begin{definition}
	Consider system \eqref{eq:appendix-dynamics} and suppose that $w\in\mc W$. 
	A function $V(\cdot):\real^n\to\real_{\geq0}$ is called an ISpS-Lyapunov function if there are some $\mc K_{\infty}$-functions $\alpha_1(\cdot),\alpha_2(\cdot)$ and $\sigma(\cdot)$, and some $\mc K$-functions $\rho_1(\cdot)$ and $\rho_2(\cdot)$ such that 
	\begin{align}
		&\alpha_1(\|x\|) \leq V(x) \leq \alpha_2(\|x\|) + \sigma(\mu) \label{eq:appendix-lyapunov-1}\\
		&V(g(x,w)) - V(x) \leq -\alpha_3(\|x\|) + \rho_1(\|w\|) + \rho_2(\mu) \label{eq:appendix-lyapunov-2}
	\end{align}
\end{definition}

\begin{lemma}\label{lemma:ISpS}
	If system \eqref{eq:appendix-dynamics} admits an ISpS-Lyapunov function, then the system is input-to-state-practically stable.
\end{lemma}

\subsection{Proof of Theorem 1}
The proof is divided into two parts. We first show recursive feasibility of the MPC optimization \eqref{eq:mpc-augmented}. 
Then we show that the system trajectory enters the set $\mc X_f$ in finite time and the parameters $C$ and $R$ are exactly estimated when $x\in\mc X_f$.
\subsubsection{Recursive feasibility}
Recursive feasibility of Algorithm~\ref{alg:adpative-mpc} requires that for any terminal sets $\mc X_f$, 
if \eqref{eq:mpc-augmented} is feasibility at time $t=0$, then it is feasible for all $t\geq0$.
We first show that for all terminal sets considered in Algorithm~\ref{alg:adpative-mpc}, there exists sufficiently large $T$ such that MPC optimization \eqref{eq:mpc-augmented} is feasible at $t=0$.
Three cases for different kinds of $\mc X_f$ in Algorithm~\ref{alg:adpative-mpc} are considered:
\begin{enumerate}
	\item \label{item:feasibility-1}
		For any $i\in\Lint, j\in\Lout_i$: 
		Consider $\mc X_u(u) = \{\bar{x},\ubar{x}\in\real^{n_x}_{\geq0}:\bar{x}_{ij}\leq \ubar{C}_{ij}S_{ij}(u), \bar{x}_{ki}\leq\ubar{C}_{ki}S_{ki}(u)\ \forall\, k\in\Lin_i\}$ for some $u\in\mc U$.
	    Fix any $k\in\Lin_i$ and for every $l\in\Lint$, 
		let $u(t)$ be such that $S_{ki}(u(t))=1$, and keep $S_{lk}(u(t))=0, S_{ij}(u(t)) = 1$. 
		This is possible by Assumption~\ref{assumption:signal}.
		Then, $\bar{x}_{ki}$ decreases and after finite time, $\bar{x}_{ki}(t+1)=\bar{x}_{ki}(t) = 0\leq\ubar{C}_{ki}S_{ki}(u)$. 
		Repeat the process by setting $S_{ki}(u(t))=1$ for each $k\in\Lin_i$, we have $\bar{x}_{ki}(t) = 0$ for all $k\in\Lin_i$.
		Then, $\bar{x}_{ij}$ decreases and after finite time, $\bar{x}_{ij}(t)\leq \ubar{C}_{ij}S_{ij}(u(t))$.
	\item 
		For any $i\in\Lint, j\in\Lout_i$: 
		Consider $\mc X_u(u) = \{\bar{x},\ubar{x}\in\real^{n_x}_{\geq0}:\ubar{x}_{ij}\geq \bar{C}_{ij}S_{ij}(u), \bar{x}_{ki}\leq\ubar{C}_{ki}S_{ki}(u)\ \forall\, k\in\Lin_i\}$ for some $u\in\mc U$.
		Following Case~\ref{item:feasibility-1}, we have $\ubar{x}_{ij}(t)\leq\bar{x}_{ij}(t)\leq \ubar{C}_{ij}S_{ij}(u(t)), \bar{x}_{ki}\leq\ubar{C}_{ki}S_{ki}(u(t))\ \forall\, k\in\Lin_i$ for some $u(t)\in\mc U$.
		Then, $\bar{x}(t)\in\mc X_u(u)$ for $u$ such that $S_{ki}(u) = S_{ki}(u(t))$ for all $k\in\Lin_i$ and $S_{ij}(u)=0$.
	\item
		For any $i\in\Lentry, j\in\Lout_i$:
		Consider $\mc X_u(u) = \{\bar{x},\ubar{x}\in\real^{n_x}_{\geq0}:\ubar{x}_{ij}\geq \bar{C}_{ij}S_{ij}(u), \bar{x}_{kj}\leq C_{kj}S_{kj}(u), \forall\, k\in\Lin_j\setminus\{i\}\}$ for some $u\in\mc U$.
		Fix any $k\in\Lin_i$ and for every $l\in\Lint$, 
		let $u(t)$ be such that $S_{lk}(u(t))=0, S_{ki}(u(t))=1$. 
		Then, following similar reasons as Case~\ref{item:feasibility-1}, 
		after finite time we have $\bar{x}_{ki}(t+1)=\bar{x}_{ki}(t) = 0\leq\ubar{C}_{ki}S_{ki}(u(t))$, and $\ubar{x}_{ij}(t)\geq\bar{C}_{ij}S_{ij}(u(t))$ with $S_{ij}(u(t))=0$.
\end{enumerate}
Suppose \eqref{eq:mpc-augmented} is feasible at time $t$, 
then there exists an optimal solution $\{\hat{u}^*(0|t),\ldots,\hat{u}^*(T-1|t)\}$ such that $(\bar{x}(T|t),\ubar{x}(T|t))\in \mc X_f$.
We now show that there exists a control $\tilde{u}\in\mc U$, such that $(F(\bar{x}(T|t),\tilde{u},\lambdaup;\bar{C},\ubar{C},\bar{R}), F(\ubar{x}(T|t),\tilde{u},\lambdalb;\ubar{C},\bar{C},\ubar{R}))\in\mc X_f$.
This will imply recursive feasibility since the sequence $\{\hat{u}^*(1|t),\ldots,\hat{u}^*(T-1|t),\tilde{u}\}$ is feasible to \eqref{eq:mpc-augmented} at $t+1$ provided that $F(x(t),\hat{u}^*(0|t),\lambdalb;\ubar{C},\bar{C},\ubar{R})\leq x(t+1) \leq F(x(t),\hat{u}^*(0|t),\lambdaup;\bar{C},\ubar{C},\bar{R})$.

For any $i\in\Lint,j\in\Lout_i$,
consider the control $\tilde{u}$ such that $S_{ij}(\tilde{u}) = S_{ki}(\tilde{u}) = S_{lk}(\tilde{u}) = 0$ for all $k\in\Lin_i,l\in\Lin_k$.
Then,  $F_{ij}(\bar{x}(T|t),\tilde{u},\lambdaup;\bar{C},\ubar{C},\bar{R}) = \bar{x}_{ij}(T|t)$, $F_{ij}(\ubar{x}(T|t),\tilde{u},\lambdalb;\ubar{C},\bar{C},\ubar{R})=\ubar{x}_{ij}(T|t)$
and $F_{ki}(\bar{x}(T|t),\tilde{u},\lambdaup;\bar{C},\ubar{C},\bar{R}) = \bar{x}_{ki}(T|t)$, $F_{ki}(\ubar{x}(T|t),\tilde{u},\lambdalb;\ubar{C},\bar{C},\ubar{R})=\ubar{x}_{ki}(T|t)$ for all $k\in\Lin_i$.
Therefore, $(F(\bar{x}(T|t),\tilde{u},\lambdaup;\bar{C},\ubar{C},\bar{R}), F(\ubar{x}(T|t),\tilde{u},\lambdalb;\ubar{C},\bar{C},\ubar{R}))\in\mc X_f$. 
The case when $i\in\Lentry$ and $j\in\Lout_i$ follows similarly.

\subsubsection{Finite termination}
To show that Algorithm~\ref{alg:adpative-mpc} terminates in finite time, 
it suffices to show that the system trajectory enters all three types of terminal sets $\mc X_f$ in finite time.
For any $\mc X_f$ considered in Algorithm~\ref{alg:adpative-mpc}, for $(\bar{x},\ubar{x})\notin\mc X_f$ with $\bar{x}\geq\ubar{x}$, we have $\bar{x}\neq0, \ubar{x}\neq0$, 
and there exists a constant $r>0$ such that $\|\bar{x}\|_1 \geq r$ since $\mc X_f$ contains an open neighborhood of the origin.
Let $l_{\min} = \min_i\{l_i:i=1,\ldots,n_x\}$, we have $\ell(\bar{x},u)\geq l_{\min}\|\bar{x}\|_1\geq l_{\min}r$ for $(\bar{x},\ubar{x})\notin\mc X_f$.
Let $V^*_{\text{aug}}(x)$ denote the value function of \eqref{eq:mpc-augmented} with $x(t)=x$.
Since $\ell(\bar{x},u)=0$ for all $(\bar{x},\ubar{x})\in\mc X_f$, we have $V^*_{\text{aug}}(x(t+1)) - V^*_{\text{aug}}(x(t))\leq - \ell(x(t),u(t))$.

Without loss of generality, suppose $(x(0),x(0))\notin\mc X_f$. 
Let $\bar{t}$ denote a finite integer such that $\bar{t}l_{\min}r>V^*_{\text{aug}}(x(0))$. 
Suppose the system trajectory has not entered $\mc X_f$ by $t=\bar{t}$,
i.e., $(x(t),x(t))\notin\mc X_f$ for all $t=0,\ldots,\bar{t}$, we have $\|x(t)\|_1\geq r$ and thus $\ell(x(t),u(t))\geq l_{\min}r$ for all $t=0,\ldots,\bar{t}$.
Then, $V^*_{\text{aug}}(x(\bar{t}))\leq V^*_{\text{aug}}(x(0)) -\bar{t}l_{\min}r<0$, which contradicts that $V^*_{\text{aug}}(\cdot)\geq0$.
Therefore, the system trajectory enters $\mc X_f$ in finite time. 
For $x\in\mc X_f$, the parameters for $C$ or $R$ are exactly estimated with formulas in Algorithm~\ref{alg:adpative-mpc} according to the dynamics \eqref{eq:dynamics}.

\subsection{Proof of Theorem 2}

\begin{proof}
	Let $\epsilon>0$ be the constant defined in Lemma~\ref{lemma:max-pressure}.
	Consider the following optimization problem \eqref{eq:mpc-optimization-lyapunov}.
	We proceed by showing that the optimal value function of \eqref{eq:mpc-optimization-lyapunov} is an ISpS-Lyapunov function for the closed-loop system \eqref{eq:dynamics} in feedback with MPC controller \eqref{eq:mpc-optimization-lyapunov}.

	Let $V(x,u)$ denote the objective value of \eqref{eq:mpc-optimization-lyapunov} with $x(t)=x,u(0|t)=u$. 
	Then, an upper bound of $V(x,u)$ can be derived as follows:
	\begin{equation}\label{eq:lyapunov-upper-bound}
		\begin{aligned}
			&V(x,u) \\
			&= \epsilon \|x\|_1 + \|\lambda\|_2^2 \\
			&\quad + \sum_{i\in\Link_{\text{entry}}}\sum_{j\in\Link^{\text{out}}_i} \big[(x_{ij} - C_{ij}S_{ij}(u))^2 + 2R_{ij}\lambda_i(x_{ij} - \bar{C}_{ij}) \\ &\hspace{2.2in}+ 2R_{ij}\lambda_i\bar{C}_{ij}\big]\\
			& \quad + \sum_{i\in\Link_{\text{int}}}\sum_{j\in\Link^{\text{out}}_i}x_{ij}^2(1|t) \\
			& \leq \epsilon \|x\|_1 \\ 
			& \quad+ \sum_{i\in\Link_{\text{entry}}}\sum_{j\in\Link^{\text{out}}_i} \big[(\max\{x_{ij} - C_{ij}S_{ij}(u),0\} + R_{ij}\lambda_i)^2 \\ & \hspace{2in}+ \bar{C}^2_{ij} + 2R_{ij}\lambda_i\bar{C}_{ij}\big]\\
			& \quad + \sum_{i\in\Link_{\text{int}}}\sum_{j\in\Link^{\text{out}}_i}x_{ij}^2(1|t) \\
			& = \epsilon \|x\|_1 + \|f(x,u,\lambda;R,C)\|_2^2  \\
			& \hspace{1.5in}+ \sum_{i\in\Link_{\text{entry}}}\sum_{j\in\Link^{\text{out}}_i} \bar{C}_{ij}^2 + 2R_{ij}\lambda_i\bar{C}_{ij}\\
		\end{aligned}
	\end{equation}
	where the inequality follows by $$(x_{ij} - C_{ij}S_{ij}(u))^2 \leq (\max\{x_{ij} - C_{ij}S_{ij}(u),0\}) ^2 + \bar{C}^2_{ij}$$ 
	and $$x_{ij} - \bar{C}_{ij}\leq\max\{x_{ij} - C_{ij}S_{ij}(u),0\}$$
	A lower bound can be similarly derived as follows:
	\begin{equation}\label{eq:lyapunov-lower-bound}
		\begin{aligned}
			&V(x,u) \\
			&\geq \epsilon \|x\|_1 \\ 
			& \quad + \sum_{i\in\Link_{\text{entry}}}\sum_{j\in\Link^{\text{out}}_i} (\max\{x_{ij} - C_{ij}S_{ij}(u),0\} + R_{ij}\lambda_i)^2\\
			& \quad + \sum_{i\in\Link_{\text{int}}}\sum_{j\in\Link^{\text{out}}_i}x_{ij}^2(1|t) \\
			& = \epsilon \|x\|_1 + \|f(x,u,\lambda;R,C)\|_2^2
		\end{aligned}
	\end{equation}
	where the inequality follows by $$\left(x_{ij} - C_{ij}S_{ij}(u)\right)^2 \geq (\max\{x_{ij} - C_{ij}S_{ij}(u),0\}) ^2$$ 
	and $$x_{ij}\geq\max\{x_{ij} - C_{ij}S_{ij}(u),0\}$$

	Let $V^*(x)$ denote the optimal value function of \eqref{eq:mpc-optimization-lyapunov}.
	We first show that \eqref{eq:appendix-lyapunov-1} holds for $V^*(x)$. 
	It is clear that $V^*(x)\geq\alpha_1(\|x\|_2) = \epsilon\|x\|_2$.
	For the upper bound, since $f(x,u,\lambda;R,C)\leq x + c$ where $c$ is the vector of nonzero entries in $C$, we have $\|f(x,u,\lambda;R,C)\|_2^2\leq \|x\|_2^2 + 2\|x\|_2\|c\|_2 + \|c\|_2^2$.
	Let $\mu$ be an upper bound on $\lambda$ such that $||\lambda||_2\leq \mu$. 
	By \eqref{eq:lyapunov-upper-bound}, $V^*(x)\leq \|x\|_2^2 + (\sqrt{n_x}\epsilon + 2\|c\|)\|x\|_2 + \sigma_2\mu$ for some sufficiently large $\sigma_2$.
	Therefore, \eqref{eq:appendix-lyapunov-1} holds for $V^*(x)$ with $\alpha_2(\|x\|_2) = \|x\|_2^2 + (\sqrt{n_x}\epsilon + 2\|c\|)\|x\|_2$ and $\sigma(\mu) = \sigma_2\mu$.

	Then, we use Lemma~\ref{lemma:max-pressure} to construct the inequality \eqref{eq:appendix-lyapunov-2} for dynamics \eqref{eq:dynamics} in feedback with MPC \eqref{eq:mpc-optimization-lyapunov}.
	Let $u^*(x)$ denote an optimal solution to \eqref{eq:mpc-optimization-lyapunov} given initial condition $x(t)=x$.
	Let $u^\text{mp}(x)$ denote the max-pressure control proposed in \cite{varaiya2013max}.
	Since $u^\text{mp}(x)$ is a feasible solution to \eqref{eq:mpc-optimization-lyapunov}, we have
	\begin{align*}
		&V^*(x(t+1)) - V^*(x(t)) \\ 
		\leq \ &V(x(t+1), u^{\text{mp}}(x(t+1))) - V(x(t),u^*(x(t))) \\
		\leq \ &\epsilon\|x(t + 1)\|_1 + \|f(x(t+1),u^{\text{mp}}(x(t+1)),\lambda;R,C)\|_2^2 \\
		& - \epsilon\|x(t)\|_1 - \|f(x(t),u^*(x(t)),\lambda;R,C)\|_2^2 \\
		& + \sum_{i\in\Link_{\text{entry}}}\sum_{j\in\Link^{\text{out}}_i} \bar{C}_{ij}^2 + 2R_{ij}\lambda_i\bar{C}_{ij} \quad \text{by \eqref{eq:lyapunov-upper-bound} and \eqref{eq:lyapunov-lower-bound}}\\
		= \ & \|f(x(t+1),u^{\text{mp}}(x(t+1)),\lambda;R,C)\|_2^2 -  \|x(t+1)\|_2^2 \\
		& + \epsilon\|x(t + 1)\|_1  -\epsilon\|x(t)\|_1  \\ 
		& \hspace{1.3in}+ \sum_{i\in\Link_{\text{entry}}}\sum_{j\in\Link^{\text{out}}_i} \bar{C}_{ij}^2 + 2R_{ij}\lambda_i\bar{C}_{ij} \\
		\leq \ & -\epsilon\|x(t)\|_1 + k + \sum_{i\in\Link_{\text{entry}}}\sum_{j\in\Link^{\text{out}}_i} \bar{C}_{ij}^2 + 2R_{ij}\lambda_i\bar{C}_{ij} \\
		&\hspace{2in} \text{by Lemma~\ref{lemma:max-pressure}}
	\end{align*}
Then \eqref{eq:appendix-lyapunov-2} holds for $V^*(x)$ with $\alpha_3(\|x\|) = \epsilon\|x\|_2$ and $\rho_2(\mu) = a\mu$ 
where $a$ is sufficiently large such that $k+\sum_{i\in\Link_{\text{entry}}}\sum_{j\in\Link^{\text{out}}_i} \bar{C}_{ij}^2 + 2R_{ij}\lambda_i\bar{C}_{ij} \leq a\mu$.
From Lemma~\ref{lemma:ISpS}, the closed-loop system \eqref{eq:dynamics} in feedback with MPC controller \eqref{eq:mpc-optimization} is input-to-state practically stable with respect to $\lambda$.
\end{proof}

\end{document}